\documentclass[10pt]{article}
\usepackage[OE]{express}

\begin{document}
\title{Maximal quantum Fisher information for phase estimation without initial parity}

\author{Xu Yu,\authormark{1} Xiang Zhao,\authormark{1,2} Luyi Shen, \authormark{1,3} Yanyan Shao, \authormark{4}
Jing Liu,\authormark{4,5} and Xiaoguang Wang,\authormark{1,6}}

\address{
\authormark{1} Zhejiang Institute of Modern Physics, Department of Physics, Zhejiang University,
Hangzhou 310027, China \\
\authormark{2} Department of Physics, ETH Zurich, 8049 Zurich, Switzerland \\
\authormark{3} Department of Engineer, Ecole Polytechnique, France \\
\authormark{4} MOE Key Laboratory of Fundamental Physical Quantities Measurement,
Hubei Key Laboratory of Gravitation and Quantum Physics, School of Physics, 
Huazhong University of Science and Technology, Wuhan 430074, China}

\email{\authormark{5}liujingphys@hust.edu.cn}
\email{\authormark{6}xgwang@zimp.zju.edu.cn}

\begin{abstract}
Mach-Zehnder interferometer is a common device in quantum phase estimation and the
photon losses in it are an important issue for achieving a high phase accuracy. Here
we thoroughly discuss the precision limit of the phase in the Mach-Zehnder interferometer
with a coherent state and a superposition of coherent states as input states.
By providing a general analytical expression of quantum Fisher information, the
phase-matching condition and optimal initial parity are given. Especially, in the photon loss
scenario, the sensitivity behaviors are analyzed and specific strategies are provided to restore the
phase accuracies for symmetric and asymmetric losses.
\end{abstract}


\section{Introduction}
Quantum metrology, an emerging quantum technology, has been widely studied~\cite{Giovannetti2011,Paris2009,
Escher2011,Tsang2013,Demkowicz2012,Demkowicz2014,Yuan2015,Yuan2016,Yuan2017,Yao2014,Toth2014,
Liu2017,Liu2017a} and applied in various sciencfic tasks in recent years, including the detection of gravitational
wave~\cite{Roman2010,Snadden1998,Chu1999}, quantum imaging~\cite{Sean2008,Ryan2008,Shih2007,Boyd2012}
and even biology science~\cite{Bowen2016}. Quantum phase estimation via interferometers is an important
topic in quantum metrology. A successful example of phase estimation with interferometers
is the Laser Interferometer Gravitational-Wave Observatory (LIGO), which has already catch the
signal of gravitational waves~\cite{LIGO2016} in 2015. Another two on-going projects LISA~\cite{Vitale2016}
and TianQin~\cite{TianQin} are also based on the orbital optical interferometers. Hence, the study of
optical phase estimation, especially quantum phase estimation, will definitely promote the technological
development in these fields, and may even breed the next-generation detectors for gravitational waves
and dark matters.

A $\mathfrak{su}(2)$ interferometer can be constructed via a Mach-Zehnder interferometer,
which typically consists of two beam splitters and one phase shift in one arm, as shown in Fig.~\ref{Fig:sketch}.
Since Caves found the effects of vacuum fluctuation to the phase accuracy in Mach-Zehnder interferometers~\cite{Caves1981},
various types of input states have been discussed, including squeezed state\cite{Caves1981,Pezze2008,Jin2017,Jin2017a},
NOON state~\cite{Wineland1994,Leibfried2004,Humphreys2013}, entangled coherent state~\cite{Joo2011,Sanders1992,
Gerry1997,Gerry2010,Joo2012,Liu2016,Israel2017}, BAT state~\cite{Cooper2010}, and number squeezed state~\cite{Pezze2013}.

A powerful theoretical tool in quantum parameter estimation to depict the precision limit is the quantum
Cram\'{e}r-Rao bound, which is $\delta\hat{\phi}\ge 1/\sqrt{\mu F}$~\cite{Helstrom1976,
Holevo1982,Braunstein1994,Giovannetti2006}. Here $\delta\hat{\phi}$ is the standard deviation of
parameter $\phi$ with unbiased estimator $\hat{\phi}$, $\mu$ is the repeated number of experiments
and $F$ is the quantum Fisher information (QFI). The most useful resource in the Mach-Zehnder
interferometer is the average photon numer $\bar{n}$, of which the corresponding standard quantum
limit for $\delta \hat{\phi}$ is $1/{\sqrt{\bar{n}}}$ and the Heisenberg limit (or Heisenberg scaling)
is $1/\bar{n}$.

Noise is the major obstacle for obtaining high precision result in quantum parameter estimation.
For a large-scale, especially an in-orbit quantum interferometer (in the size of LISA and TianQin),
the photon losses between the satellites could be an important issue for a high phase sensitivity.
Therefore, fully understanding on the sensitivity behaviors under photon losses in the interferometer
could help to restore a high precision as required. Many lossy scenarios with different input states
have been discussed in the literature~\cite{Joo2011,Joo2012,Knott2014,Takeoka2017,Demkowicz2009,Gkortsilas2012}.
It is common to simulate the photon losses with fictitious beam splitters in theory, as shown in Fig.~\ref{Fig:sketch}.
In this paper, we discuss the precision limit of a Mach-Zehnder interferometer with a coherent
state and a superposition of coherent states as the input states.
Both perfect and imperfect (with photon losses) scenarios are considered and the analytical expression of
QFI is provided. With this expression, the phase-matching condition (PMC) of the input states and the optimal
QFI are calculated. For the imperfect scenario, symmetric and asymmetric losses are both studied and corresponding
strategies to restore the accuracy are provided.

\section{Preliminary knowledge}
The quantum Fisher information (QFI) for a parameter $\phi$ is defined as
$F :=\mathrm{Tr}(\rho_{\phi} L^2)$, where $\rho_{\phi}$ is the parameterized state and
$L$ is the symmetric logarithmic derivative operator satisfying
$\partial_{\phi}\rho_{\phi}=\frac{1}{2}(L\rho_{\phi}+\rho_{\phi} L)$.
Several methods for the calculation of QFI have been developed in recent years~\cite{Zhang2013,Liu2014,Liu2014a,
Pang2014,Jiang2014,Liusr,Liu2016a}. Utilizing the spectral decomposition $\rho_{\phi}=\sum_{i=1}^{M}{p_i}|\psi_i\rangle\langle\psi_i|$,
with $M$, $p_i$ and $|\psi_i \rangle$ the dimension of the support for the matrix, the eigenvalues
and eigenstates of $\rho_{\phi}$, the QFI can be expressed by~\cite{Zhang2013,Liu2014,Liu2014a}:
\begin{equation}
F=\sum_{i=1}^{M}{\frac{(\partial_{\phi} p_i)^2}{p_i}}+\sum_{i=1}^{M}{4p_i}
\langle\partial_{\phi}\psi_i|\partial_{\phi}\psi_i\rangle
-\sum_{i,j=1}^{M}{\frac{8p_{i}p_{j}}{p_{i}+p_{j}}}|\langle\psi_{i}|\partial_{\phi}\psi_j\rangle|^2.
\label{Eq:F1}
\end{equation}
For the unitary parameterization process $\rho_{\phi}=e^{-iH\phi}\rho e^{iH\phi}$,
with $H$ a Hermitian operator, the expression of the QFI reduces to
\begin{equation}
F=\sum_{i=1}^{M}{4p_i}\langle\psi_i|H^2|\psi_i\rangle-\sum_{i,j=1}^{M}
{\frac{8 p_i p_j}{p_i+p_j}}|\langle\psi_i|H|\psi_j\rangle|^2.
\end{equation}
Furthermore, for a pure state $|\psi_{\phi}\rangle$, it reduces to
\begin{equation}
F= 4\left(\langle\psi_{\phi}|H^2|\psi_{\phi}\rangle-|\langle\psi_{\phi}|H|\psi_{\phi}\rangle|^2 \right).
\label{Eq:QFI_pure}
\end{equation}

%========================================================
\begin{figure}
\centering
\includegraphics[width=9cm]{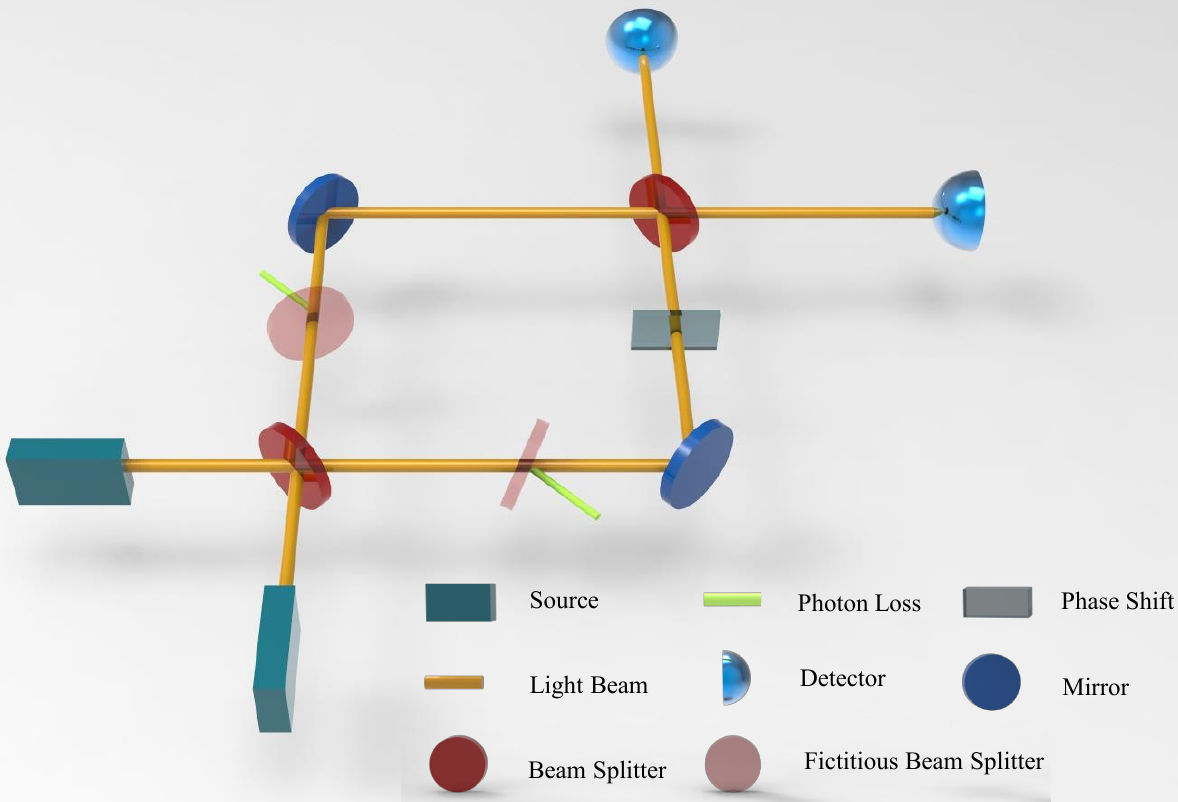}
\caption{Schematic of an Mach-Zehnder Interferometer. The input ports are labeled as A and B.
The photon losses in the interferometer are simulated with two fictitious beam splitters,
with corresponding operators $B^{T_1}_{\mathrm{AC}}$ and $B^{T_{2}}_{\mathrm{BD}}$. Here
C and D are fictitious ports and $T_{1}$, $T_{2}$ are transmission rates. No photon losses
exist for $T_{1}=T_{2}=1$.}
\label{Fig:sketch}
\end{figure}
%========================================================

In this paper, we focus on the phase estimation in the Mach-Zehnder interferometer, as shown in
Fig.~\ref{Fig:sketch}. The interferometer consists of two beam splitters and a phase shift. The two beam splitters
are usually taken as 50:50 beam splitters, which in theory can be expressed by
$B_{x}\left(\pm\frac{\pi}{2}\right)=\exp\left(\pm i\frac{\pi}{2}J^\text{AB}_x\right)$. Here $J_{x}^{\mathrm{AB}}$
is a Schwinger operator defined as $J_{x}^{\mathrm{AB}}=\frac{1}{2}(a^{\dagger}b+b^{\dagger}a)$ with $a$ ($b$) the
annihilation operator for port A (B) and $a^{\dagger}$ ($b^{\dagger}$) the corresponding creation operators.
The other two Schwinger operators are $J_{y}^{\mathrm{AB}}=\frac{1}{2i}(a^{\dagger}b-b^{\dagger}a)$ and
$J_{z}^{\mathrm{AB}}=\frac{1}{2}(a^{\dagger}a-b^{\dagger}b)$. The Schwinger operators satisfy the $\mathfrak{su}(2)$
algebra. The operator for the phase shift is $U(\phi)=\exp\left(i\phi J^{\mathrm{AB}}_z\right)$.
For a perfect Mach-Zehnder interferometer, one expression of the entire setup can be written as the
unitary operator below~\cite{Yurke1986}
\begin{equation}
U_{\mathrm{MZ}}=B_x\left(-\frac{\pi}{2}\right)U(\phi)B_x\left(\frac{\pi}{2}\right)
=\exp\left(-i\phi J^{\mathrm{AB}}_y\right).
\label{Eq:Ut}
\end{equation}

The photon loss in the interferometer is usually depicted via fictitious beam splitters in theory.
The effect of a general beam splitter for ports X and Y in quantum optics can be written as
$B^T_{\mathrm{XY}}=\exp \left(i2\arccos\sqrt{T}J_{x}^{\mathrm{XY}}\right)$~\cite{Demkowicz2009,Gkortsilas2012},
where $T$ is the transmission rate. When $T=1$ ($T=0$), all photons are transmitted (reflected).
In many scenarios, especially in the large-scale interferometers, the optical path length is long and
the dispersion of light spot is inevitable during the propagation, which will cause photon losses at the second beam splitter.
In this paper, ports A and B are input ports of the interferometer
and ports C and D are the fictitious ports for photon losses. The effects of fictitious beam splitters
are expressed by $B^{T_{1}}_{\mathrm{AC}}$ and $B^{T_{2}}_{\mathrm{BD}}$. The arm with respect to port
A (B) has no photon losses for $T_{1}=1$ ($T_{2}=1$) and all photons are lost for $T_{1}=0$ ($T_{2}=0$).

\section{Perfect interferometer}
For the perfect interferometer and with a pure input state, the QFI can be directly obtained by
substituting Eq.~(\ref{Eq:Ut}) into Eq.~(\ref{Eq:QFI_pure}), which is
\begin{equation}
F=\left(2\bar{n}_{\mathrm{A}} \bar{n}_{\mathrm{B}}+\bar{n}_\mathrm{A}
+\bar{n}_{\mathrm{B}}-2|\langle a\rangle|^2|\langle b\rangle|^2\right)\notag\\
-2\mathrm{Re}\left(\langle a^2\rangle\langle {b^\dagger}^2\rangle-\langle a\rangle^2\langle b^\dagger\rangle^2\right),
\end{equation}
where $\mathrm{Re}(\cdot)$ represent the real part and $\bar{n}_{\mathrm{A}}=\langle a^\dagger a\rangle$,
$\bar{n}_{\mathrm{B}}=\langle b^\dagger b\rangle$ are the average photon numbers for both arms.

Taking a coherent state $|\beta\rangle$ as the input state for port A, and an arbitrary pure state $|\psi\rangle$
for port B, $\bar{n}_\text A=|\langle a\rangle|^2=|\beta|^2$. The QFI then reads
\begin{equation}
F=2\bar{n}_\mathrm{A}\left(\bar{n}_{\mathrm{B}}-|\langle b\rangle|^2\right)+\bar{n}_\mathrm{A}+\bar{n}_{\mathrm{B}}
-2\mathrm{Re}\left[\beta^2\left(\langle {b^\dag}^2\rangle-\langle b^\dag\rangle^2\right)\right].
\label{Eq:Fp2}
\end{equation}
For a given $|\psi\rangle$, $F$ only depends on the value of $\beta$, and the PMC optimizing the QFI is
\begin{equation}
\left|\mathrm{Arg}\left(\beta^2\right)-\mathrm{Arg}\left(\langle b^2\rangle-\langle b\rangle^2\right)\right|=\pi,
\label{Eq:pmc1}
\end{equation}
where $\mathrm{Arg}(\cdot)$ is the argument. With this condition, the optimal QFI can be calculated as
\begin{equation}
F_{\mathrm{m}}=2\bar{n}_{\mathrm{A}}\left(\bar{n}_{\mathrm{B}}-\left|\langle b\rangle\right|^2+\left|\langle b^2\rangle
-\langle b\rangle^2\right|\right)+\bar{n}_\mathrm{A}+\bar{n}_{\mathrm{B}}.
\label{Eq:fm1}
\end{equation}

Next we take the input state in port B as the superposition of two coherent states
$|\alpha\rangle$ and $|-\alpha\rangle$, i.e.,
\begin{equation}
|\psi\rangle=N_\alpha(|\alpha\rangle+e^{i\Theta}|-\alpha\rangle),
\label{Eq:inputb}
\end{equation}
where $\Theta\in[0,2\pi)$ is the relative phase and the normalization factor $N_{\alpha}$ reads
$N_{\alpha}=(2+2e^{-2|\alpha|^2}\cos\Theta)^{-1/2}$. In the following we denote
$\beta=|\beta|e^{i\Phi_\text A}$, $\alpha=|\alpha|e^{i\Phi_\text B}$, with $\Phi_\text A$, $\Phi_\text B$
the arguments of $\beta$ and $\alpha$. Through some straightforward calculation, the PMC for optimal
QFI can be written as
\begin{equation}
\label{Eq:phi}
|\Phi_{\mathrm{A}}-\Phi_{\mathrm{B}}|=\frac{\pi}{2},
\end{equation}
which coincides with the case that using a coherent superposition state in port B~\cite{Liu2013}, namely,
the relative phase $\Theta$ doesn't affect the PMC. Under this condition, the maximal QFI reads
\begin{equation}
F_{\mathrm{m}}=\bar{n}_{\mathrm{A}}(2|\alpha|^2+1)+\bar{n}_{\mathrm{B}}(2\bar{n}_{\mathrm{A}}+1).
\label{Eq:fm2}
\end{equation}
Utilizing the equations $\bar{n}_{\mathrm{A}}=|\beta|^2$ and $\bar{n}_{\mathrm{B}}
=2N^2_{\alpha}|\alpha|^2(1-e^{-2|\alpha|^2}\cos\Theta)$, the maximal $F_{\mathrm{m}}$ can be reached
when $\Theta=\pi$, which means taking into account the PMC, the QFI can be further improved with an
initial odd parity of $|\psi\rangle$.

Now we compare $F_{\mathrm{m}}$ with Heisenberg scaling. Denote $\bar{n}=\bar{n}_{\mathrm{A}}
+\bar{n}_{\mathrm{B}}$ as the average total input photon number, Eq.~(\ref{Eq:fm2}) can then be rewritten into
$F_{\mathrm{m}}=\bar{n}+2\bar{n}_{\mathrm{A}}(\bar{n}_{\mathrm{B}}+|\alpha|^2)$.
For a large $|\alpha|$, $\bar{n}_{\mathrm{B}}\approx|\alpha|^2$,
$F_{\mathrm{m}}\approx\bar{n}+4\bar{n}_{\mathrm{A}}\bar{n}_{\mathrm{B}}=\bar{n}+\bar{n}^2-(\delta \bar{n})^2$
with $\delta \bar{n}=\bar{n}_{\mathrm{A}}-\bar{n}_{\mathrm{B}}$ the photon difference between two ports.
When $\delta \bar{n}$ is small (compared to $\bar{n}$), $F_{\mathrm{m}}$ reduces to $\bar{n}+\bar{n}^2$, i.e.,
$F_{\mathrm{m}}\propto\bar{n}^2$, indicating the QFI under PMC can reach the Heisenberg scaling even
no initial parity exists in $|\psi\rangle$. Furthermore, it can be found that
\begin{equation}
F_{\mathrm{m}}\leq \langle \hat{n}^{2}\rangle,
\end{equation}
which can be proved as $F_{\mathrm{m}}-\langle \hat{n}^2\rangle=-(|\beta|^2-|\alpha|^2)^2\le 0$. Here we used
$\langle \hat{n}^2\rangle=\bar{n}_{\mathrm{A}}^2+2\bar{n}_{\mathrm{A}}\bar{n}_{\mathrm{B}}+\bar{n}+|\alpha|^4$.
This upper bound can be achieved for $|\beta|=|\alpha|$. To satisfy this condition, one can take
$\beta=\alpha e^{i(\Phi+\frac{\pi}{2})}$ with $\Phi$ the relative phase between the values of $\alpha$ and $\beta$,
hence the total input state is $N_{\alpha}|\alpha e^{i(\Phi+\frac{\pi}{2})}\rangle_{\mathrm{A}}\otimes
(|\alpha\rangle_{\mathrm{B}}+e^{i\Theta}|-\alpha\rangle_{\mathrm{B}})$.
According to Eq.~(\ref{Eq:pmc1}), the PMC is $\Phi=0$ or $\pi$, which is indeed independent of $\Theta$.

\section{Imperfect interferometer}
For an imperfect Mach-Zehnder interferometer, the total effect cannot be treated as an unitary operation.
As discussed in the previous section, the photon losses are simulated with beam splitters
$B^{T_{1}}_{\mathrm{AC}}=\exp[i2\arccos\sqrt{T_1}J_x^\text{AC}]$ and $B^{T_{2}}_{\mathrm{BD}}=\exp[i2\arccos\sqrt{T_2}J_x^\text{BD}]$.
Here $J_x^{\mathrm{AC}}=\frac{1}{2}(a^\dagger c+c^\dagger a)$, $J_x^{\mathrm{BD}}=\frac{1}{2}(b^\dagger d+d^\dagger b)$ with $c$ ($c^\dagger$)
and $d$ ($d^\dagger$) the annihilation (creation) operators of the fictitious lossy ports C and D. We take the total input state as
\begin{equation}
N_{\alpha}|\alpha e^{i(\Phi+\frac{\pi}{2})}\rangle_{\mathrm{A}}
\left(|\alpha\rangle_{\mathrm{B}}+e^{i\Theta}|-\alpha\rangle_{\mathrm{B}}\right).
\label{Eq:input}
\end{equation}

After the photon losses, the state becomes a mixed state, which can be written as
(the basis information and detailed calculation can be found in the appendix)
\begin{equation}
\label{Eq:rho'}
\rho_{1}=N_\alpha^2\left(
\begin{array}{cc}
1+p_t^2+2e^{-2|\alpha|^2}\cos\Theta & \sqrt{1-p_t^2}\left(p_t+p_r e^{-i\Theta}\right)e^{i|\alpha|^2\delta T\sin\Phi} \\
\sqrt{1-p_t^2}\left(p_t+p_r e^{i\Theta}\right)e^{-i|\alpha|^2\delta T\sin\Phi} & 1-p_t^2 \\
\end{array}\right),
\end{equation}
where $p_t=e^{-|\alpha|^{2}T}$, $p_r=e^{-|\alpha|^{2}R}$ and $T=T_{1}+T_{2}$ is the total transmission
rate of the photon losses, $R=2-T$ is the total reflection rate, $\delta T=T_{1}-T_{2}$ is the transmission difference
between the two ports. Since the last 50:50 beam splitter in the interferometer does not affect the value
of QFI as it is independent of $\theta$, the total effect of the lossy interferometer is equivalent to
perform the phase shift transform $U(\theta)$ to $\rho_{1}$. Denote the eigenvalues and eigenstates of
$\rho_{1}$ as $\lambda_{\pm}$ and $|\lambda_{\pm}\rangle$, respectively, the QFI can be expressed by
\begin{equation}
F=\sum_{i={\pm}}{4\lambda_{i}}\langle\lambda_{i}|({J^{\mathrm{AB}}_{z}})^2|\lambda_{i}\rangle
-\sum_{i,j={\pm}}{\frac{8\lambda_{i}\lambda_{j}}{\lambda_{i}+\lambda_{j}}}|\langle\lambda_{i}|
J^{\mathrm{AB}}_{z}|\lambda_{j}\rangle|^2.
\label{Eq:F3}
\end{equation}
Utilizing the expressions of $\lambda_\pm$ and $|\lambda_\pm\rangle$ (given in the appendix) and
through some tedious calculation, the specific expression of QFI can be written as
\begin{eqnarray}
F \!&=&\! 2(\delta T)^{2}\frac{N_{\alpha}^{6}|\alpha|^{4}p_{t}}{\Delta(1-p_{t}^2)}\left[4p_{r}(p_{r}+p_{t}\cos\Theta)
(p_t+p_r\cos\Theta)- \frac{\Delta}{N_{\alpha}^{4}}p_{t}\right] \nonumber \\
& & -16(\delta T)^{2}\frac{N_{\alpha}^{8}}{\Delta}|\alpha|^{4}(1-p_r^{2})e^{-4|\alpha|^2} \sin^{2}\Theta
+2\delta T N_\alpha^{2}|\alpha|^{2}e^{-2|\alpha|^2}(4T N_\alpha^{2}|\alpha|^{2}-1)\sin\Theta\sin\Phi \nonumber \\
& & +2T^{2}|\alpha|^{4}N_{\alpha}^{2}\left[1-2N_{\alpha}^{2}(1-p_r^{2})-\left(\frac{\Delta}
{2N_{\alpha}^{2}}+2N^{2}_{\alpha}e^{-4|\alpha|^2}\sin^{2}\Theta\right)\sin^{2}\Phi\right]+2TN_{\alpha}^{2}|\alpha|^{2}.
\label{Eq:F4}
\end{eqnarray}
where $\Delta=1-4\det\rho_{1}=1-4N_{\alpha}^{4}(1-p_{r}^{2})(1-p_{t}^{2})$.
Next we will discuss the PMCs and maximum QFIs for symmetric and asymmetric losses scenarios.

\subsection{Symmetric losses}
We first consider the symmetric losses case. In this case, the transmission rate in both arms are equivalent,
i.e., $\delta T=0$. With this condition, the QFI in Eq.~(\ref{Eq:F4}) reduces to
\begin{equation}
\label{Eq:F5}
F=2T^{2}|\alpha|^{4}N_{\alpha}^{2}\left[1-2N_{\alpha}^{2}(1-p_r^{2})-\left(\frac{\Delta}
{2N_{\alpha}^{2}}+2N^{2}_{\alpha}e^{-4|\alpha|^2}\sin^{2}\Theta\right)\sin^{2}\Phi\right]+2TN_{\alpha}^{2}|\alpha|^{2}.
\end{equation}
To maximize $F$, the corresponding PMC is $\Phi=0$ or $\pi$, which is the same with lossless case.
Utilizing the PMC, $F_{\mathrm{m}}$ is in the form
\begin{equation}
F_{\mathrm{m}}=2TN_{\alpha}^{2}|\alpha|^2\left\{1+T|\alpha|^{2}\left[1-2N_{\alpha}^{2}(1-p_r^{2})\right]\right\}.
\label{Eq:Fm}
\end{equation}

%==================================================================
\begin{figure}[tp]
\centering
\includegraphics[width=8cm]{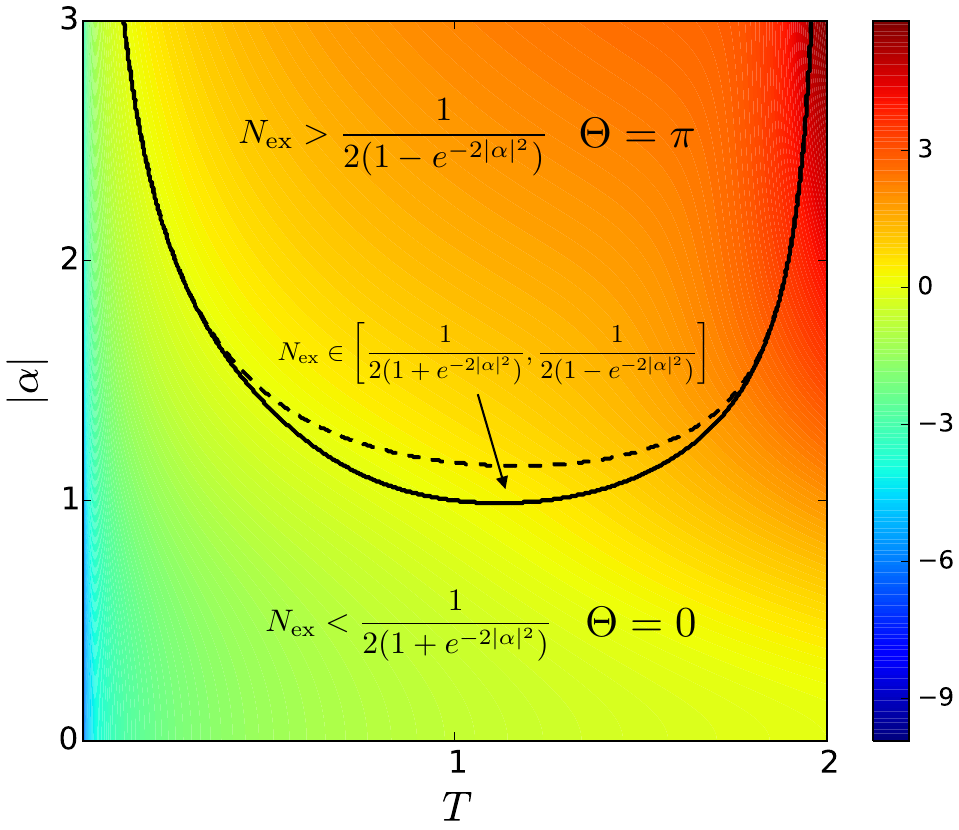}
\caption{The values of $\underset{\Theta}{\max} F_{\mathrm{m}}$ as a function of $T$ and $|\alpha|$.
The areas below the solid black line, above the dashed black line and between these lines
represent the regimes that $N_{\mathrm{ex}}<(2+2e^{-2|\alpha|^2})^{-1}$ (optimal $\Theta$ is 0),
$N_{\mathrm{ex}}>(2-2e^{-2|\alpha|^2})^{-1}$ (optimal $\Theta$ is $\pi$) and
$N_{\mathrm{ex}}\in[(2+2e^{-2|\alpha|^2})^{-1}, (2-2e^{-2|\alpha|^2})^{-1}]$.
The PMC here is $\Phi=0,\pi$. $\underset{\Theta}{\max} F_{\mathrm{m}}$ takes the logarithmic values
in the figure.}
\label{Fig:2}
\end{figure}
%==================================================================

Recall the fact that $N_{\alpha}$ is a function of $\Theta$, $F_{\mathrm{m}}$ can be further improved by
optimizing $\Theta$. Utilizing the equation $\frac{\partial F_{\mathrm{m}}}{\partial N_{\alpha}^{2}}=0$,
it can be found the extremal value of $F_{\mathrm{m}}$ is reached at
\begin{equation}
N_{\alpha}^{2}=N_{\mathrm{ex}}:=\frac{1+T|\alpha|^2}{4T|\alpha|^{2}(1-p^{2}_{r})}
\end{equation}
and due to the fact $\frac{\partial^2 F_{\mathrm{m}}}{(\partial N_{\alpha}^{2})^2}<0$, this extremal value is
the maximum value. One may notice that $N_{\alpha}^{2}$ is a bounded function with respect to $\Theta$,
\begin{equation}
N_{\alpha}^{2}\in\left[\frac{1}{2(1+e^{-2|\alpha|^2})}, \frac{1}{2(1-e^{-2|\alpha|^2})}\right],
\end{equation}
where the lower and upper bounds of $N_{\alpha}^{2}$ can be attained at $\Theta=0$ and $\pi$, respectively.
To obtain the actual maximum value of $F_{\mathrm{m}}$, whether $N_{\mathrm{ex}}$ locates in above regime needs
to be considered. The specific relation between $N_{\mathrm{ex}}$ and above regime is shown in Fig.~\ref{Fig:2}.
The areas below the solid black line and above the dashed black line represent
the regimes that $N_{\mathrm{ex}}<(2+2e^{-2|\alpha|^2})^{-1}$ and
$N_{\mathrm{ex}}>(2-2e^{-2|\alpha|^2})^{-1}$, respectively. The area between these lines represents
the regime that $N_{\mathrm{ex}}\in\left[(2+2e^{-2|\alpha|^2})^{-1}, (2-2e^{-2|\alpha|^2})^{-1}\right]$.

In the regime that $N_{\mathrm{ex}} >(2-2e^{-2|\alpha|^2})^{-1}$. $N_{\mathrm{ex}}$ is not attainable and the
maximum value of $F_{\mathrm{m}}$ with respect to $\Theta$ is obtained at $\Theta=\pi$, namely,
an initial odd parity is required. In this case,
\begin{equation}
\underset{\Theta}{\max} F_{\mathrm{m}}=\frac{T|\alpha|^2}{1-e^{-2|\alpha|^2}}\left[1+T|\alpha|^{2}
\left(1-\frac{1-e^{-2|\alpha|^2 R}}{1-e^{-2|\alpha|^2}}\right)\right].
\label{eq:symmetry1}
\end{equation}
Similarly, in the regime that $N_{\mathrm{ex}} < (2+2e^{-2|\alpha|^2})^{-1}$, $N_{\mathrm{ex}}$
is also not attainable and the maximum value of $F_{\mathrm{m}}$ with respect to $\Theta$ is obtained
at $\Theta=0$, namely, an initial even parity is required for optimal $F_{\mathrm{m}}$. In this case,
\begin{equation}
\underset{\Theta}{\max} F_{\mathrm{m}}=\frac{T|\alpha|^{2}}{1+e^{-2|\alpha|^2}}\left[1+T|\alpha|^{2}
\left(1-\frac{1-e^{-2|\alpha|^2 R}}{1+e^{-2|\alpha|^2}}\right)\right].
\label{eq:symmetry2}
\end{equation}
In the regime that $N_{\mathrm{ex}}\in[(2+2e^{-2|\alpha|^2})^{-1},(2-2e^{-2|\alpha|^2})^{-1}]$,
$N_{\mathrm{ex}}$ is reachable and the maximum $F_{\mathrm{m}}$ can be attained at
$N^{2}_{\alpha}=N_{\mathrm{ex}}$. The maximum $F_{\mathrm{m}}$ reads
\begin{equation}
\underset{\Theta}{\max} F_{\mathrm{m}}=\frac{\left(1+T|\alpha|^2\right)^2}{4(1-e^{-2|\alpha|^2 R})}.
\label{eq:symmetry3}
\end{equation}
The optimal $\Theta$ satisfies the following equation
\begin{equation}
\cos\Theta=\frac{-2T|\alpha|^2e^{-2|\alpha|^2(1-T)}+(T|\alpha|^2-1)e^{2|\alpha|^2}}{1+T|\alpha|^2}.
\end{equation}
In this regime, the optimal $\Theta$ relies on the values of $T$, $|\alpha|$ and both
odd and even input states are non-optimal.

From the lines shown in Fig.~\ref{Fig:2}, it can be seen the area between the lines
is small, which means for most values of $T$ and $|\alpha|$, $N_{\mathrm{ex}}$ is out of the regime
$\left[(2+2e^{-2|\alpha|^2})^{-1}, (2-2e^{-2|\alpha|^2})^{-1}\right]$, and
initial parity will benefit the precision limit. Besides, though the PMC here is not changed compared to
the lossless scenario, the maximum $F_{\mathrm{m}}$ with respect to $\Theta$ is different
for different parameter regimes as discussed above. However, in all regimes, increasing the intensity of
initial state always benefits the precision limit, as shown in Fig.~\ref{Fig:2}. Thus, for an intermediate
photon loss rate, the best strategy to hold the precision limit is to use a high intensity odd state as
the input state. However, for a low photon loss rate, one should be more careful since the increasing of
the intensity may requires a changed parity for optimal precision limit. And to keep the odd parity to be
optimal, a higher intensity is required with the decrease of the photon loss rate (the increase of the
transmission rate $T$).

\subsection{Asymmetric losses}
For asymmetric losses scenario, $\delta T \neq 0$. To find the PMC, the derivative of QFI on
$\sin\Phi$ needs to be calculated. Based on Eq.~(\ref{Eq:F4}), it is
\begin{equation}
\frac{\partial F}{\partial \sin\Phi}=-2\delta T N_\alpha^{2}|\alpha|^{2}e^{-2|\alpha|^2}
\sin\Theta(1-4T N_\alpha^{2}|\alpha|^{2})
-2T^{2}|\alpha|^{4}\left(\Delta+4N_{\alpha}^{4}e^{-4|\alpha|^2}\sin^{2}\Theta \right)\sin\Phi.
\end{equation}
Due to the fact $\frac{\partial^{2} F}{(\partial \sin\Phi)^{2}}<0$, the solution for above
equation gives the maximum value of QFI, i.e., the optimal $\Phi$ needs to satisfy
\begin{equation}
\sin\Phi=N^{\prime}_{\mathrm{ex}}:=\frac{N_\alpha^{2}(4T N_\alpha^{2}|\alpha|^{2}-1)\sin\Theta}
{T^{2}|\alpha|^{2}\left(\Delta e^{2|\alpha|^2}+4N_{\alpha}^{4}e^{-2|\alpha|^{2}}\sin^{2}\Theta \right)}\delta T.
\label{eq:unsmy}
\end{equation}
The solution for this equation relies on the values of $\Theta$. However, similar to the symmetric scenario,
$\sin\Phi$ is restrained in the regime $[-1,1]$. Hence, when $N^{\prime}_{\mathrm{ex}}\in[-1,1]$,
the PMC is the solution for above equation, especially, if the input state is an even state, i.e., $\Theta=0$,
the PMC then reads $\Phi=0$. For case that $N^{\prime}_{\mathrm{ex}} > 1$, the PMC is $\Phi=\pi/2$,
and for $N^{\prime}_{\mathrm{ex}} < -1$, the PMC is $\Phi=3\pi/2$.

%========================================================
\begin{figure}
\centering
\includegraphics[width=8cm]{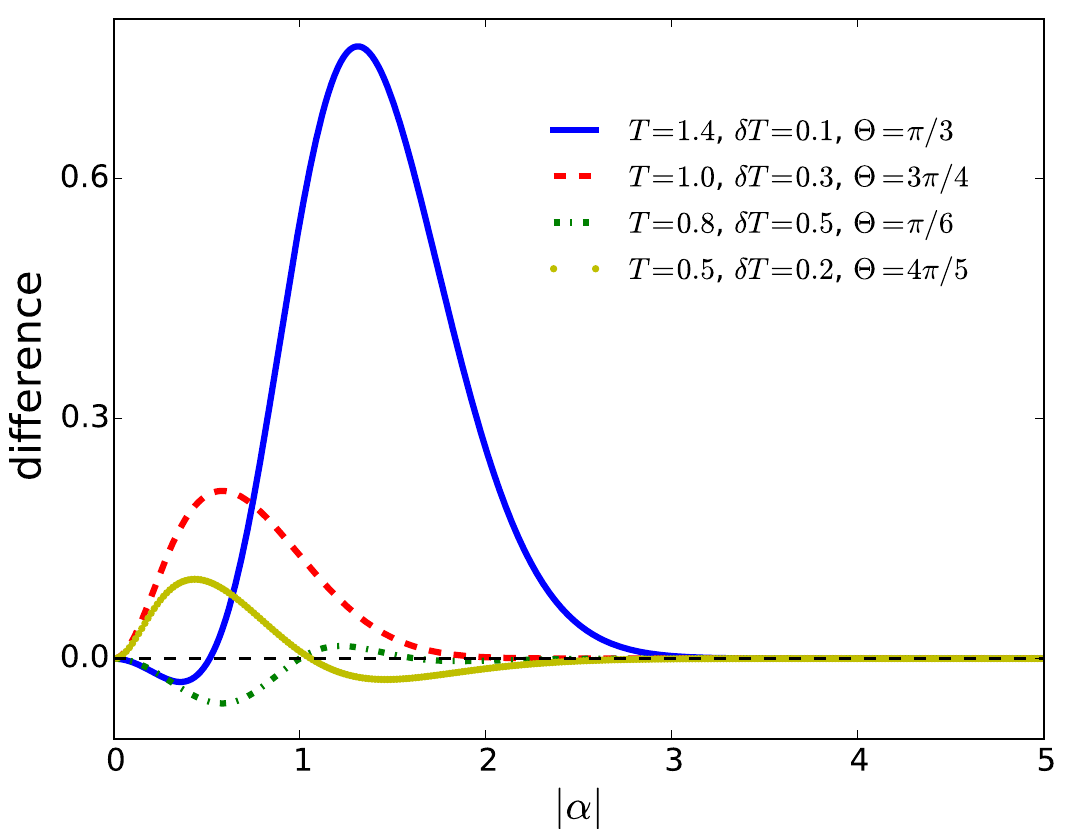}
\caption{The difference between Eq.~(\ref{Eq:F4}) and $T|\alpha|^2$ under the PMC $\Phi=0$.
The expression $T|\alpha|^{2}$ is a good approximation from $|\alpha|\approx 3$ for the coefficients values
in the figure. A larger transmission rate $T$ requires a larger $|\alpha|$ for this approximation.}
\label{Fig:approx}
\end{figure}
%========================================================

For a low intensity input ($|\alpha|$ is very small),
$N^{\prime}_{\mathrm{ex}}\approx\frac{1}{2T}\tan \frac{\Theta}{2}-\frac{\sin\Theta \delta T}
{4T^{2}|\alpha|^2}$. Its relation between the regime $[-1,1]$ highly relies on the value of $\Theta$
and the sign of $\delta T$, indicating no constant PMC exists. A more concerned case is with a high intensity input.
In this case, $N^2_{\alpha}\approx1/2$ and $\Delta e^{2|\alpha|^2} \approx e^{2|\alpha|^2(1-T)}+e^{2|\alpha|^2(1-R)}$.
Since $1-T$ and $1-R$ always take different signs as $T+R=2$, $\Delta e^{2|\alpha|^2}$
is very large here. Thus, $N^{\prime}_{\mathrm{ex}}$ approximates to zero.
Based on Eq.~(\ref{eq:unsmy}), the PMC here is $\Phi\approx 0$ or $\pi$. Namely, for an asymmetric
loss scenario with a high enough intensity input, the PMC can still be independent of $\Theta$
and the transmission rates. Another benefit of this PMC is that $F_{\mathrm{m}}$
is insensitive to the sign of $\delta T$ (since it only exists together with $\sin\Phi$),
i.e., the information of which path has a more severe leak is not required here.
Taking this PMC ($\Phi=0,\pi$), $F_{\mathrm{m}}$ approximates to
\begin{equation}
F_{\mathrm{m}} \approx T|\alpha|^{2},
\end{equation}
which is independent of $\Theta$. Figure~\ref{Fig:approx} shows the difference between
Eq.~(\ref{Eq:F4}) and $T|\alpha|^{2}$ for different parameter settings. Generally, a larger transmission
rate requires a larger $|\alpha|$ to converge to $T|\alpha|^2$.
However, for the specific parameters given in the figure, all $F_{\mathrm{m}}$ converge
before around $|\alpha|=3.0$, which means $T|\alpha|^{2}$ is a very good approximation here
for $|\alpha|> 3.0$ regardless the values of $T$, $\delta T$ and $\Theta$.
Thus, for the asymmetric losses scenario, one efficient strategy to restore the precision limit
is inputting a high intensity state satisfying the PMC.

\subsection{The scaling}
The average total photon number $\bar{n}$ for the input state in Eq.~(\ref{Eq:input}) is
\begin{equation}
\bar{n} = \frac{2|\alpha|^2}{1+e^{-2|\alpha|^2}\cos\Theta}.
\end{equation}
For a large $|\alpha|$, $\bar{n}\approx 2|\alpha|^2$. In the symmetric losses scenario,
Eqs.~(\ref{eq:symmetry1}) and (\ref{eq:symmetry2}) indicate $\underset{\Theta}{\max} F_{\mathrm{m}}
\propto \bar{n}$ for a nonzero $R$, which means it can only reach the standard quantum limit in
these regime. For Eq.~(\ref{eq:symmetry3}), $\underset{\Theta}{\max} F_{\mathrm{m}}
\propto \bar{n}^{2}$, i.e., it can still attain the Heisenberg scaling, however, the allowed value
of $|\alpha|$ in this regime (shown in Fig.~\ref{Fig:2}) is very limited, and the absolute value
of $\underset{\Theta}{\max} F_{\mathrm{m}}$ is still worse than Eq.~(\ref{eq:symmetry1}) with a
large $|\alpha|$. Therefore, the scaling of $\underset{\Theta}{\max} F_{\mathrm{m}}$ for
symmetric losses can only provide a precision at the standard quantum limit. For asymmetric losses
scenario, taking into account the PMC, $F_{\mathrm{m}}$ approximates to $T|\alpha|^2$ for high
intensity state, i.e., proportional to $\bar{n}$, the standard quantum limit.
This phenomenon coincides with some other cases that the precision limit is bounded by the standard
quantum limit when local noise exists~\cite{Escher2011,Demkowicz2012,Zhang2013,Jarzyna2015,Demkowicz2015}.

\section{Conclusion}
This paper focuses on the phase estimation of a $\mathfrak{su}(2)$ Mach-Zehnder interferometer,
in which the unknown parameter $\phi$ is encoded by a phase shift in one arm.
The input states is a coherent state $|\beta\rangle$ and a superposition state of coherent states
$N_{\alpha}(|\alpha\rangle+e^{i\Theta}|-\alpha\rangle)$. Both perfect and imperfect scenarios are considered.
For the perfect scenario, the phase-matching condition to optimize the QFI is given.
With this condition, the QFI can be further improved by taking $\Theta=\pi$, i.e., using an odd
parity state (cat state). For the imperfect scenario, the photon losses in both arms are simulated
by two fictitious beam splitters. The general analytical expression of QFI are provided, as well as the phase-matching
condition and optimal $\Theta$ to maximize QFI. In the symmetric losses case,
the phase-matching condition is unchanged compared to the lossless case. Furthermore,
there exists a small parameter regime for total transmission rate $T$ and $|\alpha|$ that optimal
$\Theta$ is sensitive to them. To avoid this regime, one strategy is using a high intensity
input state, of which the precision limit is at the standard quantum limit. However, it should be
noticed that for a large $T$, increasing the intensity may requires the change of parity from even to
odd in the mean time, and to keep the odd parity as the optimal one, a higher
intensity is required with the decrease of the photon loss rate (increase of $T$). In the asymmetric
losses case, taking the approximated phase-matching condition, an efficient strategy to avoid the
sensitivity of maximum QFI on $\Theta$ and restore the sensitivity is also using the a high intensity input state
satisfying the PMC.

\section*{Funding}
National Key Research and Development Program of China (2017YFA0205700 and 2017YFA0304202). Natural National Science Foundation (NSFC) (11475146).
Fundamental Research Funds for the Central Universities (2017FZA3005).

\section*{Disclosures}
The authors declare that there are no conflicts of interest related to this article.

\section{Appendix: Derivation of QFI for imperfect interferometer}
The input state we choose in this paper is
\begin{align}
|\psi_{\mathrm{in}}\rangle=|\alpha e^{i(\Phi+\frac{\pi}{2})}\rangle\otimes
N_\alpha(|\alpha\rangle+e^{i\Theta}|-\alpha\rangle).
\end{align}
For the first 50:50 beam splitter, the state becomes
$|\psi_{0}\rangle=B_{x}\left(\frac{\pi}{2}\right)|\psi_{\mathrm{in}}\rangle$.
Utilizing the formula
\begin{equation}
B^{T}_{x}|\alpha\rangle_{\mathrm{A}}|\beta\rangle_{\mathrm{A}^{\prime}}=\left|\alpha\sqrt{T}
+i\beta\sqrt{1-T}\right\rangle_{\mathrm{A}}\left|\beta\sqrt{T}+i\alpha\sqrt{1-T}
\right\rangle_{\mathrm{A}^{\prime}},
\end{equation}
where $B^{T}_{x}=\exp(i 2\arccos T J^\mathrm{AA^{\prime}}_x)$ and being aware of the fact
$T=1/2$ for 50:50 beam splitter, $|\psi_{0}\rangle$ can be written as
\begin{equation}
|\psi_{0}\rangle=N_\alpha\left(|if_+\rangle_{\mathrm{A}}|f_{-}\rangle_{\mathrm{B}}+e^{i\Theta}|
-if_{-}\rangle_{\mathrm{A}}|-f_{+}\rangle_{\mathrm{B}}\right),
\end{equation}
where $f_\pm:=\frac{\alpha}{\sqrt2}(1\pm e^{i\Phi})$. Recall the fictitious beam splitters
$B^{T_1}_{\mathrm{AC}}$, $B^{T_1}_{\mathrm{BD}}$ as
$\exp(i2\arccos\sqrt{T_1}J^{\mathrm{AC}}_x)$ and $\exp(i2\arccos\sqrt{T_2}J^{\mathrm{BD}}_x)$,
where C, D are labels of two fictitious output ports with $c$ ($c^\dag$) and $d$ ($d^\dag$) the annihilation (creation)
operators and $J^{\mathrm{AC}}_x=\frac{1}{2}(a^\dagger c+ac^\dagger)$, $J^{\mathrm{BD}}_x=\frac{1}{2}(b^\dagger d+bd^\dagger)$.
Assume the input states of the fictitious input ports are vacuum, and after the photon losses, the output state
$|\psi_{1}\rangle$ can be written as
\begin{equation}
|\psi_{1}\rangle=N_{\alpha}\left(|\mathcal{A}\rangle|-f_{+}\sqrt{R_1}\rangle_{\mathrm{C}}|if_{-}\sqrt{R_2}
\rangle_{\mathrm{D}}+e^{i\Theta}|\mathcal{B}\rangle|f_{-}\sqrt{R_1}\rangle_{\mathrm{C}}|-if_+\sqrt{R_2}
\rangle_{\mathrm{D}}\right),
\end{equation}
where $|\mathcal{A}\rangle:=|if_+\sqrt{T_1}\rangle_{\mathrm{A}}|f_{-}\sqrt{T_2}\rangle_{\mathrm{B}}$ and
$|\mathcal{B}\rangle:=|-if_-\sqrt{T_1}\rangle_{\mathrm{A}}|-f_{+}\sqrt{T_2}\rangle_{\mathrm{B}}$.
$R_1=1-T_1$, $R_2=1-T_2$ are the reflection rates. The reduced matrix can then be calculated as
\begin{eqnarray}
\rho_{1} \!\!&=&\!\! \mathrm{Tr}_{\mathrm{CD}}(|\psi_{1}\rangle\langle\psi_{1}|) \notag\\
\!\!&=&\!\! N^{2}_{\alpha}\left[|\mathcal{A}\rangle\langle\mathcal{A}|\!+\!|\mathcal{B}\rangle\langle\mathcal{B}|
\!+\!p_{r}e^{i(|\alpha|^2 \delta T\sin\Phi-\Theta)}|\mathcal{A}\rangle\langle\mathcal{B}|\!+\!p_{r}
e^{-i(|\alpha|^2 \delta T\sin\Phi-\Theta)}|\mathcal{B}\rangle\langle\mathcal{A}|\right]\!\!,
\end{eqnarray}
where $p_{r}=e^{-|\alpha|^{2}R}$ with $\delta T=T_1-T_2$
the transmission difference. Notice $|\mathcal{A}\rangle$ and $|\mathcal{B}\rangle$ are not orthogonal due to
the fact $\langle\mathcal{A}|\mathcal{B}\rangle=p_{t}e^{i|\alpha|^2 \delta T\sin\Phi}$ with
$p_{t}=e^{-|\alpha|^{2}T}$ and $T=T_{1}+T_{2}$ the total transmission rate.

Now we introduce an orthogonal basis $\{|\mathcal{A}\rangle,|\mathcal{A}_{\bot}\rangle\}$, where
\begin{equation}
|\mathcal{A}_{\bot}\rangle=\frac{1}{\sqrt{1-p^2_t}}\left(|\mathcal{B}\rangle-p_t
e^{i|\alpha|^2 \delta T\sin\Phi}|\mathcal{A}\rangle\right).
\end{equation}
In this basis, $\rho_{1}$ can be written as
\begin{equation}
\rho_{1}=N_\alpha^2\left(
\begin{array}{cc}
1+p_t^2+2e^{-2|\alpha|^2}\cos\Theta & \sqrt{1-p_t^2}\left(p_t+p_r e^{-i\Theta}\right)e^{i|\alpha|^2\delta T\sin\Phi} \\
\sqrt{1-p_t^2}\left(p_t+p_r e^{i\Theta}\right)e^{-i|\alpha|^2\delta T\sin\Phi} & 1-p_t^2 \\
\end{array}\right),
\end{equation}
The eigenvalues for this matrix are $\lambda_{\pm}=N^{2}_{\alpha}(1\pm\sqrt{\Delta})/2$ and corresponding eigenstates
$|\lambda_{\pm}\rangle$ are
\begin{equation}
|\lambda_\pm\rangle = \left(v_\pm \frac{p_t+p_re^{-i\Theta}}{\sqrt{p_{t}^2+p_{r}^2+2e^{-2|\alpha|^2}\cos\Theta}}
\mp\frac{p_t v_\mp }{\sqrt{1-p_t^2}}\right)|\mathcal{A}\rangle\pm \frac{v_\mp e^{-i|\alpha|^2\delta T\sin\Phi}}
{\sqrt{1-p_t^2}}|\mathcal{B}\rangle,
\end{equation}
where we have used the expression of $|\mathcal{A}_{\bot}\rangle$. The coefficients read
\begin{eqnarray}
v_\pm &=&\frac{1}{\sqrt{2}}\sqrt{1\pm \frac{p_{t}^{2}+e^{-2|\alpha|^2}\cos\Theta}
{\sqrt{\Delta}(1+e^{-2|\alpha|^2}\cos\Theta)}},\\
&=&\frac{1}{\sqrt{\Delta}}\sqrt{\frac{\Delta}{2}\pm \sqrt{\Delta}N^{2}_{\alpha}
(p_{t}^{2}+e^{-2|\alpha|^2}\cos\Theta)},\\
\Delta &=& 1-4\det\rho_{1}.
\end{eqnarray}

\section*{Acknowledgments}
The authors thank X. Xiao and J. Chen for helpful discussions.

\end{document}